# Tortuosity Measurement and the Effects of Finite Pulse Widths on Xenon Gas Diffusion NMR Studies of Porous Media

R. W. Mair*, M. D. Hürlimann**, P. N. Sen**, L. M. Schwartz**,
S. Patz***, and R. L. Walsworth*

\* Harvard-Smithsonian Center for Astrophysics, Cambridge, MA 02138, USA.
\*\* Schlumberger-Doll Research, Ridgefield, CT 06877, USA.
\*\*\* Dept. of Radiology, Brigham and Women's Hospital, Boston, MA 02115, USA

**Corresponding Author:**
Ross Mair
Harvard Smithsonian Center for Astrophysics,
60 Garden St, MS 59,
Cambridge, MA, 02138,
USA
Phone: 1-617-495 7218
Fax: 1-617-496 7690
Email: rmair@cfa.harvard.edu

# ABSTRACT

We have extended the utility of NMR as a technique to probe porous media structure over length scales of ~ 100 - 2000 μm by using the spin 1/2 noble gas $^{129}$Xe imbibed into the system's pore space. Such length scales are much greater than can be probed with NMR diffusion studies of water-saturated porous media. We utilized Pulsed Gradient Spin Echo NMR measurements of the time-dependent diffusion coefficient, $D(t)$ of the xenon gas filling the pore space to study further the measurements of both the surface area-pore volume ratio, $S/V_p$, and the tortuosity (pore connectivity) of the medium. In uniform-size glass bead packs, we observed $D(t)$ decreasing with increasing $t$, reaching an observed asymptote of ~ 0.62 - 0.65$D_0$, that could be measured over diffusion distances extending over multiple bead diameters. Measurements of $D(t)/D_0$ at differing gas pressures showed this tortuosity limit was not affected by changing the characteristic diffusion length of the spins during the diffusion encoding gradient pulse. This was not the case at the short time limit, where $D(t)/D_0$ was noticeably affected by the gas pressure in the sample. Increasing the gas pressure, and hence reducing $D_0$ and the diffusion during the gradient pulse served to reduce the previously observed deviation of $D(t)/D_0$ from the $S/V_p$ relation. The Pade approximation is used to interpolate between the long and short time limits in $D(t)$. While the short time $D(t)$ point lay above the interpolation line in the case of small beads, due to diffusion during the gradient pulse on the order of the pore size, it was also noted that the experimental $D(t)$ data fell below the Pade line in the case of large beads, most likely due to finite size effects.

# INTRODUCTION

NMR techniques are well known as non-invasive methods for study of a wide variety of porous materials. By detecting the $^1$H signal from water-saturated rocks and model systems, information including the surface-area-to-volume ratio ($S/V_p$) [1,2], average pore size [3,4] and visualizations of fluid transport under flow [5] can be obtained. Usually, these methods study the diffusion of the water spins, and the Pulsed Gradient Spin Echo (PGSE) technique has become a powerful tool for this purpose [6,7]. Except in cases of very small pores, however, spin relaxation quenches the NMR signal before water molecules can diffuse across even one pore. For this reason, liquid phase NMR has been unable to yield measurements of the long-range properties of most porous media; e.g., tortuosity in reservoir rock samples [2,8] - an important parameter that describes pore connectivity and fluid transport though the rock.

We have recently shown that similar diffusion NMR techniques can be applied to porous media imbibed with a gas rather than a liquid, yielding the time-dependent gas diffusion coefficient $D(t)$ [9,10]. The spin $\frac{1}{2}$ noble gases ($^3$He and $^{129}$Xe) are particularly ideal for such studies, given their high diffusion coefficients, their inert nature, their low surface interactions which reduce surface $T_1$ effects, and the ability to tailor the desired diffusion coefficient to some extent by altering the gas pressure in the sample. Using $^{129}$Xe gas as the observation spin, with a diffusion coefficient ~ 3 orders of magnitude higher than that of water ($5.7 \times 10^{-6}$ m$^2$ s$^{-1}$ at 1 bar pressure [9]), it has been possible to extend the porous length scales over which diffusion is observed by over an order of magnitude. We have shown that pore length scales on the order of millimeters, rather than tens of micrometers, can be probed; and that NMR gas diffusion can be used for tortuosity determination in reservoir rocks [10].

Before 1994, gas phase NMR had been a largely neglected field, due to the belief that little useful information could be obtained from a sample whose density was ~ 1000 times lower than water. The advent of the spin-exchange optical pumping technique to produce large, non-equilibrium spin polarizations (~10%) in samples of noble gas effectively cancelled the spin density deficit, allowing gas phase samples to be created with equivalent magnetization density to water in fields ~ 1 T [11]. As a result, the technique has become popular in medical imaging, especially as a non-invasive and high-resolution probe of the human and animal lung space [12]. We believe that gas phase NMR also has a vital role to play in materials science investigations, and in particular, that the advent of laser-polarized gases has increased the awareness of the potential for gas phase NMR in general.

Whether the gas is laser or thermally polarized, similar information can often be derived from NMR experiments on porous samples, although the technique and time-scale for obtaining the data may differ. As such, we believe both laser-polarized and thermally polarized gases offer the potential for complementary information from different types of samples in the wide field of porous media study. We have therefore shown that laser-polarized xenon provides identical $D(t)$ information in model porous media to that derived from thermally polarized samples [10]. However, efficient gas polarization and handling facilities are required for laser-polarized Xe NMR to provide data faster than by signal averaging from a thermal sample. As a result, while developing such a facility, we have used thermally polarized xenon to further probe some of the peculiarities of time-dependent gas diffusion NMR in porous media.

One of the key issues that arises from the use of gas-phase diffusion measurements in porous media is that of the high gas diffusion coefficients themselves. While this feature of the gases permits the probing of long length scales, it also means the gas has the potential to diffuse distances that are significant in relation to the enclosing pore space during the application of the diffusion-encoding gradient pulse in the PGSE sequences. Such a phenomenon violates a basic assumption in the PGSE experiment, i.e., the narrow pulse approximation that assumes the gradient pulse is infinitely narrow such that the spins do not diffuse during its application [7]. We believe this breakdown of the narrow pulse approximation led to deviations in gas time-dependent diffusion coefficients, $D(t)$, from values expected from the $S/V_p$ of model packed bead samples [10]. In this paper, we report the first measurements of $D(t)$ as a function of diffusion length during the gradient pulse, and show that reducing the gas diffusion coefficient can indeed reduce the previously seen deviations of $D(t)$ from the theoretical values. In addition, we apply the Pade approximation method, previously used successfully to interpolate between the short and long time water $D(t)$ limits [1,2], in order to assess the influence of excess gas diffusion during the diffusion encoding on the observed gas $D(t)$ data in the short and intermediate $t$ range.

## MATERIALS AND METHODS

*Theory*
It is well known that the NMR echo signal observed in a PGSE experiment has a Fourier relationship to the probability of spin motion – the so-called displacement propagator, which can be thought of as a spectrum of motion. The echo signal, $E$, obtained in a PGSE experiment can thus be written as [7]:

$$E(\mathbf{q},t) = \int \overline{P}_s(\mathbf{R},t) \times \exp[i2\pi\mathbf{q}\cdot(\mathbf{R})]d\mathbf{R} \qquad (1)$$

where $\overline{P}_s(\mathbf{R}, t)$ is the ensemble average displacement propagator, or the probability of a spin having a displacement $\mathbf{R} = \mathbf{r'} - \mathbf{r}$ proceeding from any initial position $\mathbf{r}$ to a final position $\mathbf{r'}$ during the 'diffusion time' $t$ (often referred to as $\Delta$ in the literature). $\mathbf{q}$ is the wavevector of the magnetization modulation induced in the spins by a field gradient pulse of strength $g$ and pulse duration $\delta$. The magnitude of $\mathbf{q}$ is $\gamma\delta g/2\pi$, where $\gamma$ is the spin gyromagnetic ratio. In the limit of small **q**, $\overline{P}_s$ is a gaussian, and the echo will be attenuated by a factor $\exp(-4\pi^2 q^2 D(t) t)$ [7], where $D(t) = \langle [\mathbf{r'} - \mathbf{r}]^2 \rangle / 6t$ is the time-dependent diffusion coefficient describing incoherent random motion of the spins in the pore space. $D(t)$ will vary as a function of $t$, with more spins encountering barriers to their motion as $t$ is increased.

For short diffusion times (i.e., small $t$), the fraction of fluid spins whose motion is restricted by pore boundaries is $\sim (S/V_p)\sqrt{D_0 t}$, where $S/V_p$ is the pore surface-area-to-volume-ratio and $\sqrt{D_0 t}$ is the characteristic free-spin-diffusion length scale for a diffusion time $t$. Using this basic physical picture, Mitra et al. have calculated, for small $t$ [13,14]:

$$\frac{D(t)}{D_0} = 1 - \frac{4}{9\sqrt{\pi}} \frac{S}{V_p} \sqrt{D_0 t} + O(D_0 t) \quad (2)$$

This important relation has been verified experimentally with NMR of liquids imbibed in a variety of model porous media [1], with the knowledge that for random bead packs, $S/V_p = 6(1-\phi)/(\phi b)$ where $\phi$ is the sample's porosity and $b$ the bead diameter. If the effects of enhanced relaxation at pore boundaries can be neglected, then for large $t$, $D(t)$ approaches an asymptotic limit of $1/\alpha$ according to [15]:

$$\frac{D(t)}{D_0} = \frac{1}{\alpha} + \frac{(1/\alpha - 1)\theta}{t} + O\left(\frac{1}{t}\right)^{\frac{3}{2}} \quad (3)$$

where $\alpha$ is the tortuosity of the porous medium [16], and $\theta$ is a fitting parameter that has units of time and appears to scale with bead size [1,2]. In dense random beads packs $1/\alpha \sim \sqrt{\phi}$ [17]. This limit has been observed in bead packs and rocks only from gas-phase time dependent diffusion measurements [10]. The Pade approximation has been successfully used in the past to extrapolate between these two limits [1,2], and this method is used here with gas-phase diffusion data for the first time:

$$\frac{D(t)}{D_0} = 1 - (1 - \frac{1}{\alpha}) \times \frac{(4\sqrt{D_0 t}/9\sqrt{\pi})(S/V_p) + (1 - 1/\alpha)(D_0 t/D_0 \theta)}{(1 - 1/\alpha) + (4\sqrt{D_0 t}/9\sqrt{\pi})(S/V_p) + (1 - 1/\alpha)(D_0 t/D_0 \theta)} \quad (4)$$

*NMR technique*

In practice, the very short spin coherence time ($T_2$) of fluid spins in a porous sample, and the high background gradients that result from susceptibility contrast between the solid grains and the imbibed fluid [18] make the simple spin echo technique unsuitable for measuring $D(t)$. Instead, we used a modified stimulated echo sequence incorporating alternating bi-polar diffusion encoding gradient pulses which serve to cancel out the effect of the background gradients while applying the diffusion encoding gradient pulses [19,20]. The sequence is illustrated in Fig. 1, where the timing parameters labeled correspond to the description in the previous section. The additional 180° pulses around each bi-polar pulse pair serve as additional refocusing pulses to help overcome fast inhomogeneous signal loss, while the crusher gradients, applied in an orthogonal axis, remove spurious signal resulting from imperfect 180° pulse calibration and $B_1$ inhomogeneity. The gradient pulses are half-sine shaped in order to reduce eddy current ring-down after the application of the gradient pulse. $D(t)$ was obtained from the resulting data by fitting the natural log of the measured echo attenuation to a modified form of the Stejskal-Tanner equation [20]:

$$\ln(S(g,t)/S(0,t)) = -g^2\gamma^2\delta^2(2/\pi)^2 D(t)(t - \delta/8 - T/6) \qquad (5)$$

where all timing parameters are described above, and the $(2/\pi)^2$ factor accounts for the half-sine shape of the gradient pulses. This equation is only applicable in the limit of $q > 1/L$, where $L$ is the sample length [2]. This limit is breached for low strength gradient pulses. As a result, the value for $g = 0$ was omitted from all fits, and the remaining gradient values had to be chosen to be high enough to avoid this regime. In addition, gradient values had to be kept low enough so as not to leave the gaussian, low-$q$ limit, which was characterized by the plot of $\ln(S(g,t)/S(0,t))$ vs $g^2$ deviating from linear, which usually occurred around a maximum signal attenuation of ~ 20%, as has been previously noted [2].

The bi-polar stimulated echo technique (PGSTE-bp) was implemented on a GE Omega/CSI spectrometer (GE NMR Instruments, Fremont, CA), interfaced to a 4.7 T magnet with clear bore of 20 cm. Experiments were performed at 55.348 MHz for [129]Xe using a tuned home-made solenoid NMR RF coil. Applied magnetic field gradients up to 7 G/cm in strength were available. In these measurements, $\delta$ was fixed at 750 μs in all experiments, the minimum possible without distortion of the gradient shape on this system. $t$ was varied from a minimum of 25 ms to a maximum of 3 s. Values of $g$ were chosen to produce significant attenuation of the [129]Xe signal while remaining between the high and low $q$ regimes described above. The combined parameters resulted in $T = 4$ ms in all experiments.

*Thermally Polarized Xenon Gas NMR experiments*
Samples of random packed spherical glass beads of different sample diameters were prepared for thermally polarized xenon experiments. Cylindrical glass cells of volume ~ 50 cm$^3$ held the bead-packs. Each cell contained beads of a single size, with bead diameters ranging from 0.1 mm to 4 mm. Appropriate amounts of xenon (isotopically enriched to 90% [129]Xe) and oxygen were frozen inside the cells at liquid nitrogen temperature. The cells were then sealed and warmed to room temperature, allowing the gases to evaporate and establish the desired imbibed-gas partial pressures inside the cell. Typically, this was 3 bar for xenon and 2 bar for oxygen, although some experiments were performed with ~ 6 bar xenon and ~ 1.5 bar oxygen. Paramagnetic oxygen was included with the xenon in order to reduce the [129]Xe $T_1$ from tens of seconds to ~ 1.3 - 1.5 s, thereby enabling efficient signal averaging which is essential due to the low NMR signal expected with thermally polarized [129]Xe gas. The PGSTE-bp sequence was employed on all samples, with 12 different $g$ values used, a repetition rate of ~ 7-8 s, and the number of signal averaging scans ranging from 16 - 256, depending on sample linewidth, signal strength and diffusion time $t$ used.

# RESULTS AND DISCUSSION

A number of static samples, each containing a different size of beads and 3 bar pressure of xenon gas, were used to obtain Xe $D(t)$ measurements across a wide range of bead sizes and diffusion lengths using the thermal xenon polarization created in the NMR magnet. This data is shown in Fig. 2, some of which first appeared in ref [10], but is reproduced here to allow comparison with data from smaller beads. The data is plotted as the reduced diffusion coefficient, $D(t)/D_0$, as a function of normalized diffusion length, $b^{-1}\sqrt{D_0 t}$, where $D_0$ is the free gas diffusion coefficient and $b$ is the bead diameter. The two characteristic limits for short and long $t$ (Eqs. (2 - 3)) are shown on the figure in straight lines, The Pade approximation has been used to interpolate between the long and short $t$ limits and hence show the expected trend of $D(t)$ between the two limits, as has been successfully demonstrated previously for water $D(t)$ in bead pack samples [1]. However, for this particular data display, the Pade fitting parameter $\theta$ has simply been adjusted to provide a fit with both the short-$t$ and long-$t$ limits over the longest diffusion distance possible. In other words, we interpolate the limiting behavior at short and long times using Eq. (4), rather than attempting to fit the line through the experimental data itself.

With appropriate signal averaging, thermally polarized xenon $D(t)$ data can show the initial short-$t$ decrease of $D(t)$, related to $S/V_p$, and can also provide an easy measure of the inverse tortuosity, $1/\alpha$, which for these samples is $1/\alpha \sim 0.63 \sim \sqrt{\phi}$, implying $\phi \sim 0.397$ is the sample porosity. The line of the short-$t$ limit is calculated using this observed $\phi$: $S/V_p = 6(1-\phi)/(\phi b) = 9.11/b$ m$^{-1}$. The data points for all beads, with the exception of the 0.1 mm beads, approach and reach the expected long-$t$ limit of $1/\alpha = 0.63$. For the largest beads, the data also lies on the calculated line of the short-$t$ limit, however, as the bead size decreases, there is an increasing deviation between the experimental points and the short-$t$ limit, as well with the medium-$t$ Pade line.

The power of this technique for studying heterogeneous structures is shown most clearly by the large normalized diffusion lengths that are observed in the smaller bead samples, which have dimensions close to those encountered in natural porous media, As shown in Fig. 2b, the diffusion lengths for this data indicate that $^{129}$Xe atoms can diffuse many bead diameters before spin depolarization limits the NMR measurement, thus probing very long distance sample heterogeneity and permitting an easy and accurate measurement of inverse tortuosity. Such multi-pore diffusion cannot be measured with NMR for liquids imbibed in porous media with pores > 50 μm [1,2,8], and, we believe, are the first measurements of diffusion over such multiple-diameter length scales in model systems, unaided by flow. (The limiting length from water diffusion arises from the water $T_1$ of ~ 1 sec and the diffusion coefficient of ~ $2 \times 10^{-9}$ m$^2$ s$^{-1}$, giving a maximum diffusion length of ~ $\sqrt{D_0 T_1} \approx \sqrt{2 \times 10^{-9}} \approx 45$ μm.)

The one sample that does not display the expected long-time asymptote for $D(t)/D_0$ is the pack of 0.1 mm glass beads. This sample exhibits a minimum reduced $D(t)$ of ~ 0.7, implying a greater porosity than for packs made from larger beads, and disagreeing with calculations of $\phi$ based on a dense, random pack of non-penetrating but otherwise non-interacting spheres [17]. The larger porosity of packed samples made from these 0.1 mm beads was verified independently by helium porosimetry, an invasive technique, and is likely a result of clumping of the very small beads.

Aside from the successful determination of sample tortuosity, the important feature of the plots in Fig. 2 is the deviation of the experimental data from the calculated short-$t$ limit derived from the measured porosity of the sample, and from the expected medium-$t$ behaviour. The most obvious trend is that, except for the largest beads, short-$t$ data generally lies above the dashed $S/V_p$ line, and that this deviation gets worse as bead size becomes smaller, down to 0.5 mm. In the 0.1 mm beads, the internal void sizes are so small as to prevent a successful measure of any short-$t$ decrease in $D(t)$, given the high diffusion coefficient of the gas which permits the Xe to sample most of the pore space within the minimum diffusion time $t$. A subtle effect is also notable in the larger beads, where the medium-$t$ $D(t)$ data initially drops below the expected Pade behavior for $D(t)$, and appears to reach the tortuosity limit earlier than is predicted by this interpolation routine. As the bead size drops below 1 mm, however, both the short and medium-$t$ data move above the expected $S/V_p$ and Pade lines, and then drop down to follow this line towards the tortuosity limit at ~ 0.5 bead diameters.

We believe this deviation of the experimental data in the larger bead samples below the Pade approximation line may be linked to the finite size of the sample. With the sample container having an inner diameter ~ 4 cm, the sample can have, at most, a width of 10 bead diameters in the case of the 4 mm beads, and 20 beads in the case of the 2 mm beads. Such conditions are likely to impose significant boundary restrictions on the sample, and invalidate the assumption of an infinite sample. The fact that this deviation reduces as the bead size is decreased gives credence to this hypothesis. This matter will be the subject of future simulations and experimental studies.

The deviation of the short-$t$ data above the calculated $S/V_p$ line as the bead size is reduced has been discussed previously [10]. We believe this deviation is most likely due to the xenon atoms diffusing a significant fraction of the pore size during the application of the diffusion encoding gradient pulse, i.e., $\sqrt{D_0 \delta} \approx a$, where $a$ is the characteristic pore length. Such an occurrence is a violation of the narrow pulse approximation, a central simplification in the analysis of diffusion experiments obtained from the PGSE technique. While the effects of this violation have been studied theoretically for echo attenuation data at high gradient strength [21,22] it has been difficult to approach this regime experimentally, given the inherent diffusion coefficient of water and other commonly used fluids.

Taking the measured free gas diffusion coefficient, $D_0$, for the xenon-oxygen mixture used in these experiments as $1.36 \times 10^{-6}$ m$^2$ s$^{-1}$ [9], the characteristic diffusion distance during the spectrometer's minimum gradient pulse time of 750 μs is approximately 32 μm. Additionally, the internal "spherical" pore space in the voids between packed beads is approximately 1/4 the bead diameter. Therefore, the 1 and 0.5 mm beads can be thought of as having $a \sim$ 250 and 125 μm "spherical pores" with narrower necks running off them. Therefore, if the gas is diffusing ~ 32 μm during the diffusion encoding gradient pulse, the spins may be covering ~ 13 - 26% of the "spherical pore" diameter, and an even greater percentage of the pore space in the narrower neck regions. Recent calculations by Wang et al. [23] suggest that the finite pulse approximation is valid when this diffusion distance during the gradient pulse is less than about 14% of the pore diameter (i.e., $0.14a$). Thus, for pores $\geq$ 200 μm, we conclude that time-dependent diffusion measurements of imbibed $^{129}$Xe gas at reasonable pressures of a few atmospheres provide a powerful probe of both small- and large-scale structure in porous media (i.e., both $S/V_p$ and tortuosity). In samples with smaller pores, however, this technique may give misleading results if used to measure $S/V_p$, because of the breakdown of the narrow pulse approximation.

With the minimum gradient pulse time a function of the spectrometer hardware, and not easily changed, the simplest way to alter the effect of $\sqrt{D_0\delta}$ was to change the gas pressure, remembering that $D_0$ scales inversely with gas pressure. The glass sample cell containing the 0.5 mm bead pack was thus filled with ~ 6.6 bar enriched $^{129}$Xe , and ~ 1.5 bar O$_2$. A new sample cell design was used for these higher pressure measurements which included a second chamber, separated from the glass beads by a porous glass frit. This design allowed the second chamber to equilibrate at the same final gas pressure as the bead cell, and thus allowed an internal check of the final Xe gas pressure by measuring $D_0$ in situ. Fig. 3 shows the $D(t)$ measurements from the 0.5 mm bead pack at both pressures of ~ 3 and ~ 6.6 bar Xe, corresponding to a Xe $D_0$ reduction from an expected value of $1.36 \times 10^{-6}$ m$^2$ s$^{-1}$ to a measured value of $0.79 \times 10^{-6}$ m$^2$ s$^{-1}$. This change in $D_0$ in turn reduced $\sqrt{D_0\delta}$ from ~ 32 to 24 μm, only a modest change; however a clear effect is seen on the resultant $D(t)$ measured at short times with the new gas pressure. The data points move down closer to the Pade line, and at the shortest times, the $S/V_p$ line. (Here, the $S/V_p$ line was calculated from the porosity that provided the best fit to the asymptotic long-$t$ $D(t)$ values.) For this data, $1/\alpha \sim 0.65 = \sqrt{\phi}$, $\phi = 0.422$ and the expected $S/V_p = 6(1-\phi)/(\phi b) = 8.22/b$ m$^{-1}$. The data clearly indicate that diffusion during the gradient pulse on a length scale approaching that of the characteristic pore length is a significant contributor to the observed deviations. This is an important constraint on gas diffusion NMR to be borne in mind, especially when more rapidly diffusing $^3$He is employed. Interestingly, while $\delta$ does not change throughout any of these experiments, and hence $\sqrt{D_0\delta}$ is constant, it appears the diffusion during the gradient pulse does not effect long-$t$ $D(t)$ measurements, and so it may be safely assumed that low pressure gases, or gases such as laser-polarized $^3$He may be suitable for tortuosity measurements in a variety of samples, even if $S/V_p$ cannot be accurately determined.

## CONCLUSIONS

Studies of time-dependent gas diffusion coefficients by NMR are proving a powerful new probe of porous media. In this current work, we have extended the normalized diffusion distance over which $D(t)$ is probed in model porous media, and have shown that the long-distance heterogeneity can be easily probed by observing $D(t)$ over diffusion distances exceeding 10 bead diameters in small bead (0.1 mm) model porous media that are representative of naturally occurring samples. Thus, similar normalized diffusion distances can be probed in model systems to those we have studied previously in reservoir rock samples [10]. The ability to measure $D(t)$ over distances of so many bead or grain diameters permits unambiguous and accurate determination of the inverse tortuosity in such samples. In addition, the ability to easily observe the tortuosity limit in model systems gives the potential for future studies of the approach to the $D(t)$ limit from the intermediate $t$ regime. $D(t)$ in the Indiana Limestone rock sample studied previously [10] showed a long, slow, non-monotonic decrease towards the eventual tortuosity limit. The ability to observe similar distances in model systems will allow the study of this $D(t)$ decrease in heterogeneous model systems and possibly the determination of multiple length scales from the $D(t)$ plot in such samples.

The diffusion studies as a function of pressure in the 0.5 mm bead pack have shown that the long-$t$ $D(t)$ measurements, from which inverse tortuosity is derived, are unaffected by varying gas pressure in the sample, and hence by the diffusion of spins during the application of the diffusion-encoding gradient pulse, even though this characteristic diffusion length may be significant on the pore length scale. This may permit accurate inverse tortuosity determination using gas phase $D(t)$ measurements with a variety of gases, such as with laser-polarized $^3$He or gas at low pressures, despite high characteristic diffusion lengths during the gradient pulse. Conversely, the short-$t$ $D(t)/D_0$ is effected by the characteristic diffusion length of the spins in the pores during diffusion encoding. An increase in gas pressure, or reduction in $D_0$ in the sample clearly influences the value of $D(t)/D_0$ at short time, and serves to reduce the deviation between the expected and observed values in this regime. This consideration will be important in any study of gas-phase diffusion in a restricted system where quantitative information is desired, and specifically in this case, has the ability to effect any such derived $S/V_p$ measurement of the enclosing sample. In addition, we have used the Pade approximation to interpolate between the short and long-$t$ limits of $D(t)$, and observed that the experimental data does not always agree with the interpolation either, due to both diffusion during the gradient pulse causing the data to rise above the Pade line as bead sizes get smaller, and possibly finite-size effects of the sample which results in the data falling below the interpolated line for large bead samples.

## ACKNOWLEDGEMENTS

We gratefully acknowledge scientific discussions with Prof. David Cory at MIT. This work was supported by NSF grant BES-9612237, NASA grants NAGW-5025 and NAG5-4920, the Whitaker Foundation, and the Smithsonian Institution Scholarly Studies Program.

# FIGURE CAPTIONS

Fig. 1. Pulse sequence diagram for the Pulsed Gradient Stimulated Echo with alternating bi-polar gradient pulses (PGSTE-bp), as used in this work. The diffusion encoding gradient pulses, of length $\delta$ and strength $g$, are shown in gray, while crusher gradients are shown in black. The diffusion time is denoted $t$, and the total diffusion encoding time is $T$. See text for further description.

Fig. 2. Time-dependent diffusion coefficient measurements for thermally polarized xenon gas imbibed in samples of randomly packed spherical glass beads. Each sample contains beads of a uniform diameter. The $^{129}$Xe time-dependent diffusion coefficient is normalized to the free gas diffusion coefficient, $D_0$, and plotted versus normalized diffusion length, $b^{-1}\sqrt{D_0 t}$. The calculated limits at short-$t$ ($S/V_p$) and long-$t$ (tortuosity) are shown by the dashed and solid lines respectively, while the solid curved line represents the interpolation between these two limits using the Pade approximation. (a) Xenon diffusion in packs of 1, 2, 3 and 4 mm diameter beads. Some of this data appeared in ref. [10]. (b) Long-$t$ (i.e., long-range) xenon gas diffusion for packs of 0.1, 0.5 and 1 mm diameter beads, showing diffusion over distances exceeding many bead diameters. Representative error bars are plotted for the 1 mm (a) and 0.1 mm (b) bead samples.

Fig. 3. $^{129}$Xe time-dependent diffusion coefficient measurements for two different xenon gas pressures in a sample of randomly packed 0.5 mm glass beads. As in Fig. 2, $D(t)/D_0$ is plotted against $b^{-1}\sqrt{D_0 t}$, and the short-$t$ and long-$t$ limits, and the Pade approximation are shown by the dashed, solid and curved lines respectively. As the gas pressure is increased, there is a noticeable reduction in the deviation between the experimental points and theoretical lines at short diffusion times. Error bars are shown for the high-pressure Xe $D(t)$ data, but are usually the size of the plot symbol.

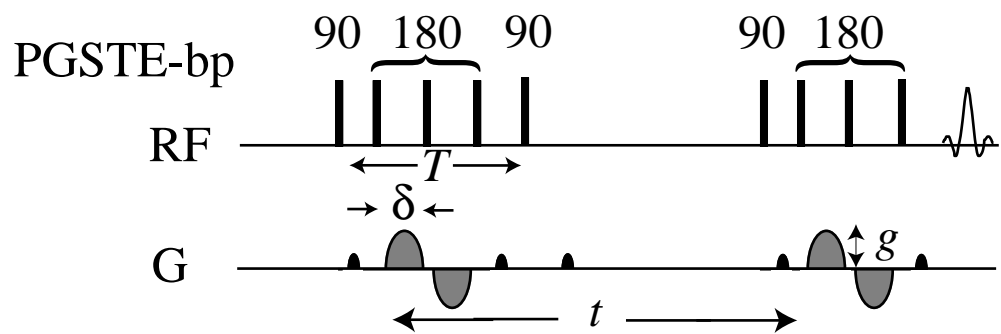

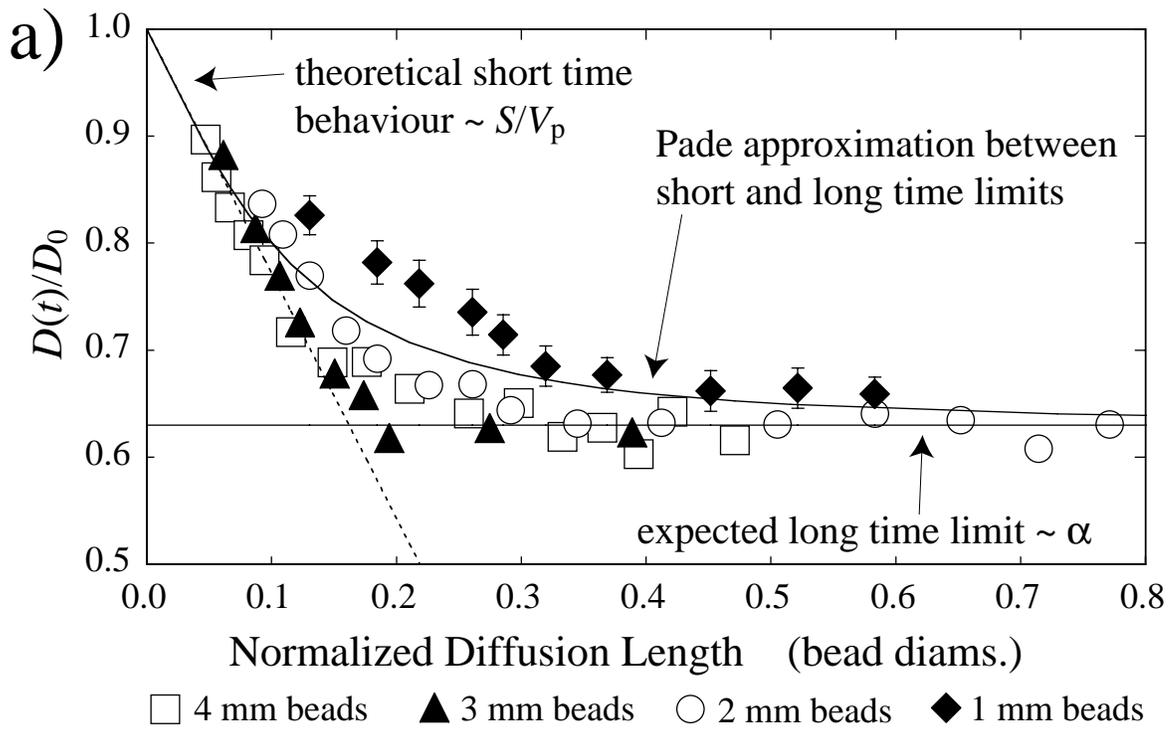

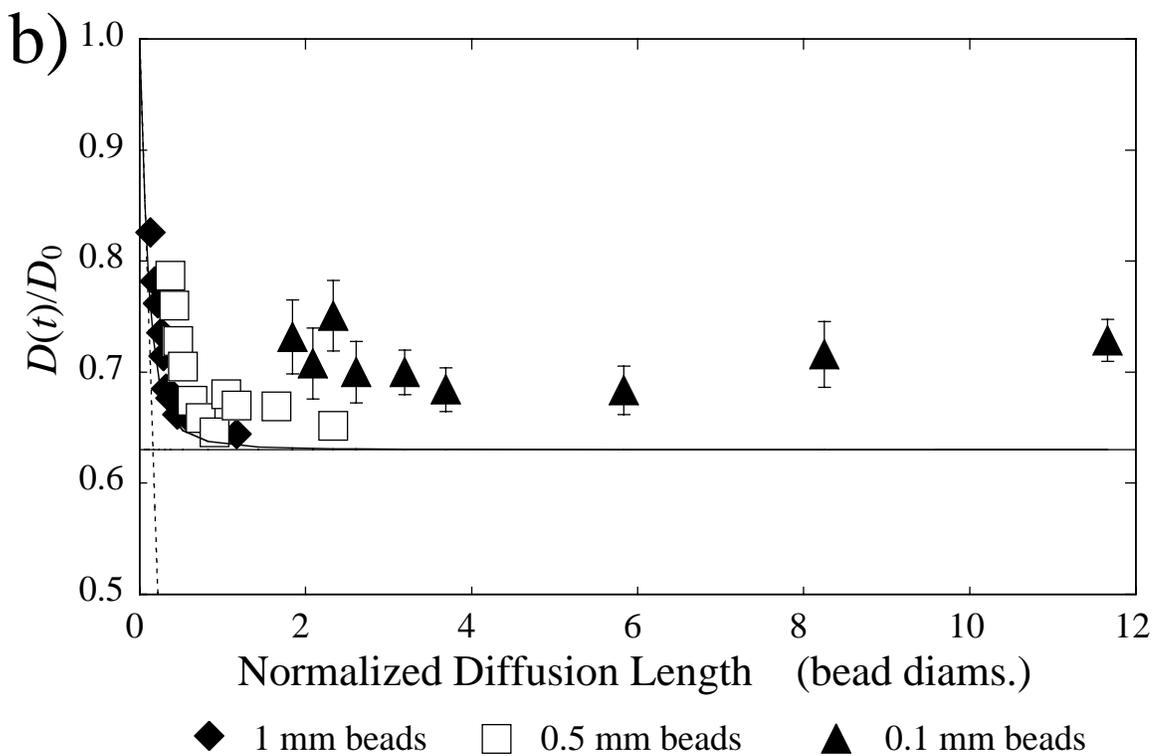

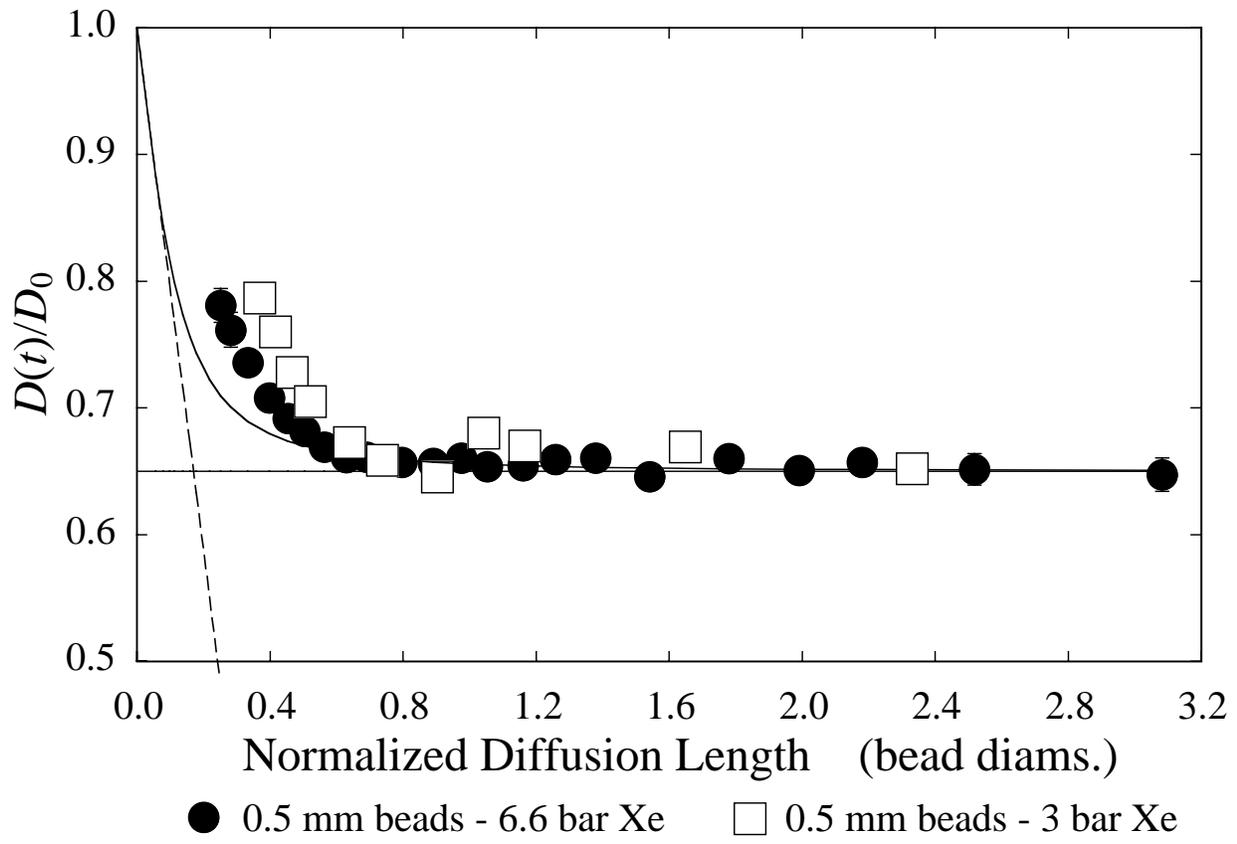